\documentclass{article}

\usepackage{graphicx}
\usepackage[dvipsnames]{xcolor}
\usepackage{booktabs}
\usepackage{subcaption}
\usepackage{array}
\usepackage{hyperref}
\usepackage{authblk}

\usepackage{mathptmx}
\usepackage{soul}\setuldepth{article}

\def\hb{\hbox to 11.5 cm{}}

\title{A Community Survey on SHACL and ShEx:\\
{\Large Briding Gaps in RDF Validation}}
\author[a]{Maxime Jakubowski}
\author[a,b]{Dominik Tomaszuk}
\author[a]{Katja Hose}
\date{}
\affil[a]{\normalsize TU Wien \\ \small \texttt{firstname.lastname@tuwien.ac.at}}
\affil[b]{\normalsize University of Bialystok \\ \small \texttt{d.lastname@uwb.edu.pl}}

\begin{document}
\maketitle

\begin{abstract}
  This paper examines RDF validation practices and challenges to
  understand stakeholder applications, their needs, and identify areas
  for improvement in technologies and methodologies, thereby guiding
  future research and standardization efforts.  A community survey was
  conducted, targeting a diverse group of RDF validation technology
  users across academia and industry. The survey collected data on
  current practices, tool usage, perceived benefits, limitations, and
  desired enhancements to gain a broad overview of the validation
  landscape.  Our analysis shows that while RDF validation is widely
  adopted and valued for enhancing data quality, significant
  challenges remain. In particular, users report a need for better
  documentation, improved tool support, enhanced performance, and
  greater language expressiveness to handle complex large-scale
  validation tasks effectively.  This work provides crucial insights
  into the RDF validation landscape, highlighting current practices
  and key areas for development. It offers a foundation for
  researchers, developers, and standardization bodies to address
  current limitations and advance validation technologies, ultimately
  improving data quality and usability in knowledge graphs.
\end{abstract}

\section{Introduction}
Knowledge Graphs (KGs) have become a ubiquitous data representation
paradigm used across the public Web, and within numerous
organizational contexts~\cite{kg_industry}. Their inherent flexibility
and broad applicability across diverse domains have driven this
widespread adoption. Recently, its relevance is further established by
the growing importance in grounding AI systems in facts, using
techniques like Graph Retrieval-Augmented Generation (GraphRAG) in
relation to
LLMs~\cite{peng2024graph,zhang2025survey,edge2024localglobal}. Prominent
examples of publicly available KGs include
WikiData~\cite{vrandevcic2014wikidata}, YAGO~\cite{suchanek2024yago},
and DBpedia~\cite{dbpedia}. Many of these public KGs leverage the
Resource Description Framework (RDF)~\cite{rdf11concepts} as their
underlying storage model, which offers advantages such as a flexible
and schemaless approach to data modeling.

As the KG landscape matured and the volume of RDF data grew, the need
for mechanisms to enforce structural constraints and ensure data
quality became increasingly apparent. This requirement led to the
development of so-called \emph{shape languages}, which provide a means
to define \emph{shapes} that express structural constraints on RDF
graphs. These shapes enable the specification of expected patterns and
properties within the graph data. The two most prominent shape
languages are the Shapes Constraint Language (SHACL)~\cite{shacl} and
Shape Expressions (ShEx)~\cite{shex}, which have garnered significant
attention and adoption within the community. While SHACL and ShEx
present syntactic differences, the primary function of these
technologies is \emph{validation}, i.e., checking whether a given RDF
graph adheres to the defined shapes.

Since the World Wide Web Consortium's (W3C) recommendation of SHACL in
2017, interest and adoption of shape languages have grown steadily in
both practical applications and academic research. This increasing
relevance is evidenced by the continually expanding body of related
literature and, notably, the recent establishment of a new W3C working
group dedicated to developing the next version of SHACL (SHACL
1.2)~\cite{datashapes_wg}.

Despite this clear momentum and ongoing evolution of standards, a
deeper understanding of the current RDF validation landscape from the
community's perspective is needed. Specifically, detailed insights are
often lacking regarding which aspects of shape languages like SHACL
and ShEx are most valued or challenging in practice, how validation
techniques are being applied across diverse contexts (particularly in
industry), and what primary obstacles hinder wider adoption or more
effective use. Therefore, this paper aims to investigate these
practical applications, perceived community needs, and challenges
related to RDF validation. Understanding these nuances is crucial not
only for informing standardization efforts but also for helping guide
future research towards real-world requirements and bridging
identified gaps. To gather this information, we conducted a
comprehensive community survey targeting a diverse audience across
both industry and academic settings involved with RDF validation.

This paper is structured as follows. Section~\ref{sec:structure}
discusses essential details about the survey’s structure and data
collection. While Section~\ref{sec:insights} presents insights into
the current state of the RDF validation landscape,
Section~\ref{sec:challenges} highlights the needs and opportunities
for further research and development. Then, Section~\ref{sec:sql}
presents our open dataset, how to reproduce our charts and tables, and
how to derrive your own insights. Section~\ref{sec:conclusion}
concludes the paper, distilling overarching insights and implications
for both practitioners and researchers.


\section{Survey Design and Data Collection} 
\label{sec:structure}

To gain insights into current practices, challenges, and community
needs surrounding RDF validation technologies (primarily SHACL and
ShEx), we conducted a comprehensive online survey between December
2024 and March 2025, yielding 94 responses. The survey utilized a mix
of multiple-choice and free-text questions, allowing us to gather both
quantitative data and richer qualitative perspectives. Outreach
efforts used relevant technical mailing lists and social media
platforms to reach interested practitioners and researchers,
supplemented by direct invitations to active SHACL/ShEx researchers
identified via the DBLP SPARQL endpoint\footnote{The query can be
  found here: \url{https://sparql.dblp.org/ux1k67}.}.

The survey investigated several key areas to build a holistic picture:
respondents' backgrounds and experience levels; the contexts and
domains where validation is applied; specifics of validation usage
including tooling, data scale, perceived benefits, and limitations;
adoption patterns for specific SHACL features; methodologies used for
shape creation; and desired future developments or
improvements. Several questions were adapted from the earlier work by
Rabbani et al.~\cite{rabbani2022shacl}, enabling comparison and
providing insights into how practices, particularly around shape
creation, may have evolved since 2022.

Throughout this paper, specific findings are sometimes linked to the
survey question numbers for clarity. The survey can still be accessed
online\footnote{\url{https://forms.gle/LdXsG644obcgSsAE6}}
for reference and is also still open to collect future responses.

\paragraph{Supplementary Material} The data from the survey can be
found on
GitHub\footnote{\url{https://github.com/dmki-tuwien/rdf-validation-community-survey/}}. All
relevant graphs, data needed to reproduce the figures from this paper,
as well as further details on the survey methodology, can be
found on our
website\footnote{\url{https://dmki-tuwien.github.io/rdf-validation-community-survey/}}. We
also exposed the survey data as a DuckDB SQL database that can be
queried in a
browser\footnote{\url{https://dmki-tuwien.github.io/rdf-validation-community-survey/databaseshell}}.

\section{Validation Landscape: Insights and Trends}
\label{sec:insights}
In this section, we will address three main themes extracted from the
survey data.

\begin{figure}
    \centering 

    \begin{subfigure}[b]{0.48\linewidth}
        \centering
        \includegraphics[width=\linewidth]{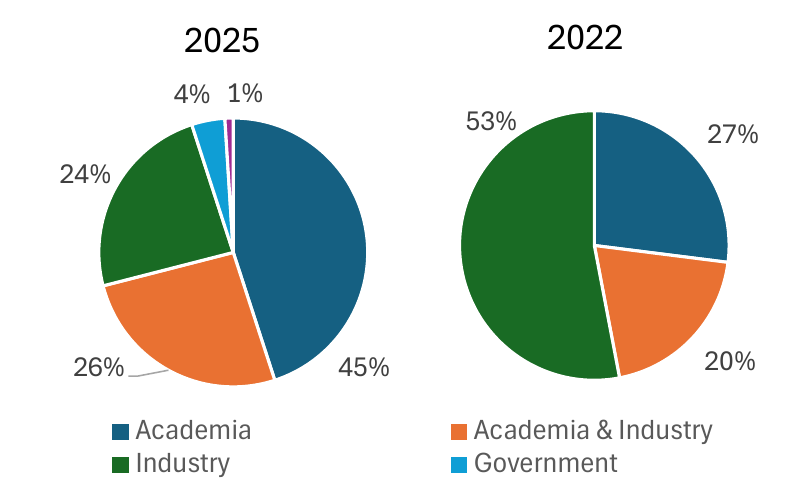}
        \caption{Professional Background}
        \label{fig:professional_background}
    \end{subfigure}
    \begin{subfigure}[b]{0.48\linewidth}
        \centering
        \includegraphics[width=\linewidth]{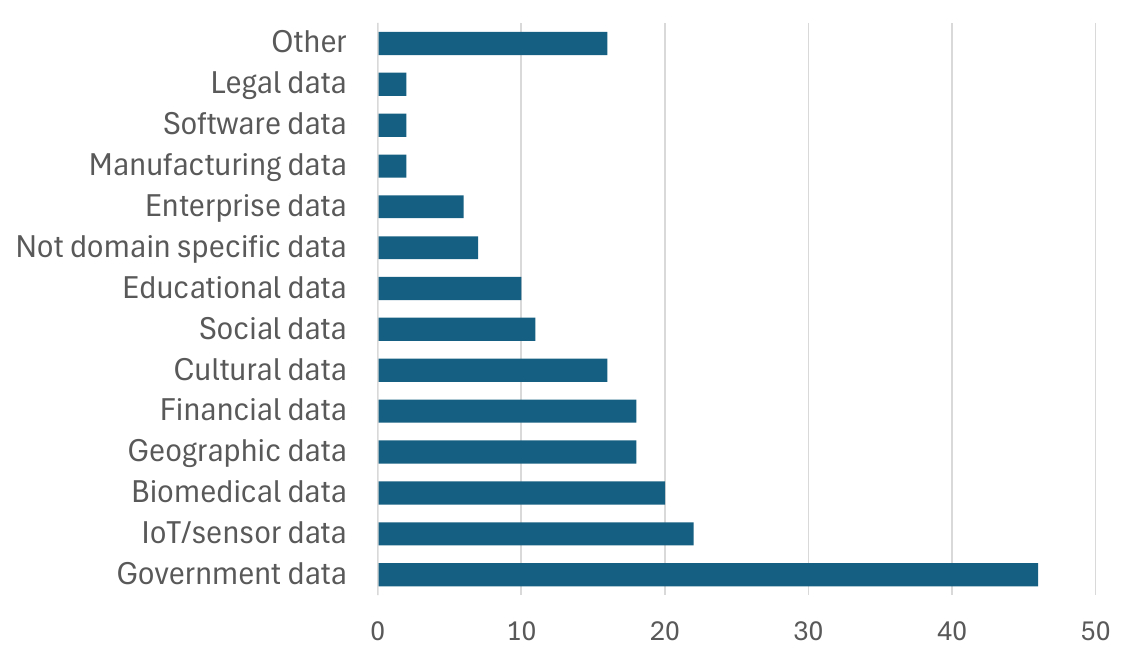}
        \caption{Application Domains}
        \label{fig:domains}
    \end{subfigure}
    
    \begin{subfigure}[b]{0.48\linewidth}
        \centering
        \includegraphics[width=\linewidth]{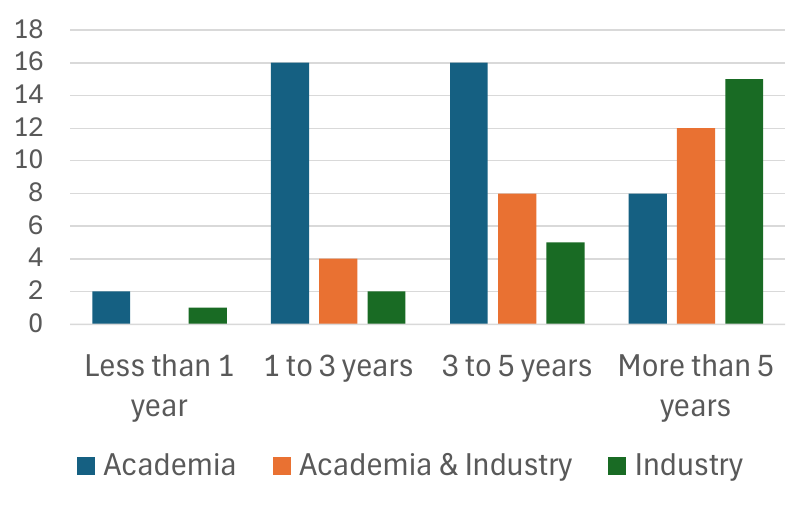}
        \caption{Years of Experience}
        \label{fig:yearsexperience}
    \end{subfigure}
    \begin{subfigure}[b]{0.48\linewidth}
        \centering
        \includegraphics[width=\linewidth]{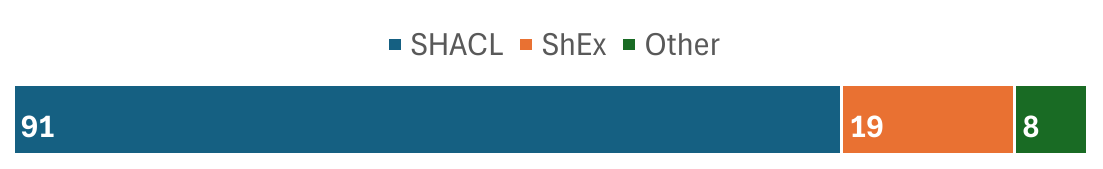}
        \caption{Language Familiarity}
        \label{fig:shapelanguages}
    \end{subfigure}

    \caption{User Background and Usage Profile} 
    \label{fig:profilingtheuser} 
\end{figure}

\paragraph{Profiling the users.}
The survey data allows us to profile the community engaging with RDF
validation, shedding light on \textbf{who the users are and the
  contexts in which these technologies are applied}. A key aspect of
the user profile is the growing scale and establishment of the user
base, evidenced by the strong response to the current survey (94
respondents) compared to the 30 respondents in the 2022
survey~\cite{rabbani2022shacl}. Figure~\ref{fig:professional_background}
reveals a notable growth in participation from the academic community
compared to 2022. Additionally, Figure~\ref{fig:yearsexperience}
suggests that adoption within academic circles has particularly
accelerated in the last one to three years, likely influenced in part
by the maturity and standardization of SHACL since its recommendation
in 2017. Nevertheless, we also observe a strong industry
presence. Additionally, there is notable usage within government
contexts, with government data being one of the most frequently cited
domains (Figure~\ref{fig:domains}), potentially reflecting drivers
like open data initiatives or data quality mandates. Generally, we see
that validation technologies are used over a varied number of domains,
highlighting their broad applicability.

Another key characteristic revealed in the user profile concerns
experience with different validation languages. Among the available
shape languages, SHACL is clearly the most widely used within
the surveyed community, with the vast majority of respondents
indicating familiarity (Figure~\ref{fig:shapelanguages}). This pattern
of widespread experience with SHACL appears consistent across
different professional backgrounds observed in the survey. While we
cannot pinpoint the exact reasons for this extensive familiarity using
the survey data, it is likely influenced significantly by SHACL's W3C
recommendation status, alongside factors like perceived technical
advantages or available tooling.

In summary, the user profile emerging from the survey indicates that
validation technologies are maturing beyond early adopters,
establishing a strong foothold in industry while gaining significant
traction within academia. Their application spans diverse domains,
with notable government usage suggesting potential roles in ensuring
data quality and compliance. While SHACL is the dominant language,
likely boosted by standardization and familiarity, ShEx maintains a
relevant user base.

\begin{figure}
    \centering
    \includegraphics[width=0.7\linewidth]{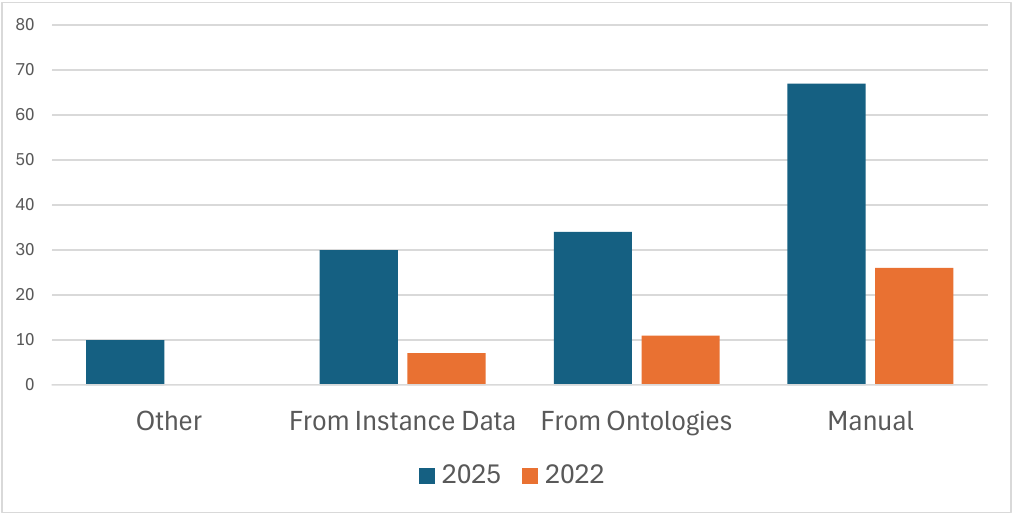}
    \caption{Methods for Shape Creation}
    \label{fig:q22_historical}
\end{figure}

\paragraph{Shape Creation Practices.}
Our survey also provides insights into the \textbf{common practices
  surrounding shape creation}, revealing the methods and tools
currently in use. As can be seen in Figure~\ref{fig:q22_historical},
manual authoring remains the predominant method. However, automated
generation techniques are also widely employed. Compared to findings
from 2022~\cite{rabbani2022shacl}, there is a notable increase in
generating shapes directly from instance data.
Interestingly, this holds when looking accross
backgrounds, and levels of experience, indicating that this is a
general trend.

Free-text responses also highlighted the practice of generating
validation shapes from diverse non-RDF artifacts, most frequently
citing sources like JSON Schema~\cite{pezoa2016foundations} and
LinkML~\cite{moxon2021linked}. This suggests a need to bridge RDF
validation with other schema specifications or higher-level conceptual
models. This is an area with potential for new tooling and
investigation, which currently appears under-explored. Furthermore,
respondents explicitly noted a need for clearer mappings between OWL
ontologies and SHACL shapes.

Regarding shape authoring tooling (Table~\ref{tab:q23_tools}), general
text editors, Protégé, and commercial solutions like TopBraid EDG are
the most common tools reported for authoring shapes. Conversely, while
there has been active research~\cite{shexer,rabbani2023}, the
landscape for \emph{tools} specifically targeting shape extraction
(e.g., from data) appears more fragmented. This may indicate a less
mature or more diverse tooling ecosystem for this specific task
compared to general shape authoring.

In summary, while \emph{manual authoring persists, automated methods,
  particularly data-driven shape generation, are gaining ground}. The
practice of generating shapes from various non-RDF artifacts and the
limited state of shape extraction tooling represent key opportunities
for future research and development.

\paragraph{Validation Practices and Tooling.} The survey leads to
valuable insights into the \textbf{operational aspects of RDF
  validation}. Looking at the validation tooling
(Table~\ref{tab:shacl_tool_usage}), the survey reveals a diverse
ecosystem where tool prominence is \emph{significantly influenced by
  underlying programming languages and their established library
  support}. For instance, widely adopted SHACL solutions like Apache
Jena (Java) and PySHACL (Python) as well as popular ShEx tools like
PyShEx and Apache Jena ShEx, are tied to these mainstream ecosystems,
likely facilitating their integration. Beyond these, the significant
presence of commercially supported tools like TopBraid EDG
(TopQuadrant) and rdf-validate-shacl (Zazuko), especially in industry,
underscores the influence of dedicated vendor support on tool
selection.

\begin{table}
  \begin{subtable}[t]{0.4\textwidth} 
    \centering
    \footnotesize
    \begin{tabular}{lrr}
      \toprule
      \textbf{Generator/Extraction} & \textbf{2025} & \textbf{2022} \\
      \midrule
      Astrea~\cite{astrea}                 &  1 &  2 \\
      Protégé~\cite{protege}                & 10 &  6 \\
      RDFShape~\cite{rdfshape}               &  1 &  5 \\
      SHACL Play~\cite{shaclplay}             &  5 &  1 \\
      SHACLGEN~\cite{shaclgen}               &  2 &  1 \\
      Shape Designer~\cite{shapedesigner}         &  1 &  1 \\
      SheXer~\cite{shexer}                 &  0 &  3 \\
      Spahiu et al.~\cite{spahiu}         &  1 &  1 \\ 
      Text Editor            & 39 &  8 \\
      TopBraid EDG~\cite{topbraid}  &  5 & 10 \\
      UnSHACLed~\cite{unshacled}              &  1 &  1 \\
      \bottomrule
    \end{tabular}
    \caption{Shape Generation/Extraction}
    \label{tab:q23_tools}
  \end{subtable}
  \hfill 
  \begin{subtable}[t]{0.59\textwidth}
    \centering
    \footnotesize
    \begin{tabular}{lrrrr}
      \toprule
      \textbf{Validator} & \textbf{Academia} & \textbf{Industry} & \textbf{Both} \\ 
      \midrule
      Apache Jena SHACL~\cite{jenashacl}      & 16 & 12 & 12 \\
      PySHACL~\cite{pyshacl}                & 17 &  8 & 13 \\
      TopBraid EDG~\cite{topbraid}           & 10 &  9 &  6 \\
      RDF4J~\cite{rdf4j}                  & 10 &  6 &  3 \\
      rdf-validate-shacl~\cite{rdfvalidateshacl}     &  1 &  5 &  3 \\
      shacl-engine~\cite{shaclengine}           &  3 &  2 &  4 \\
      shacl-js~\cite{shacljs}               &  4 &  0 &  4 \\
      maplib~\cite{maplib}                 &  1 &  2 &  0 \\
      Corese SHACL~\cite{coreseshacl}           &  2 &  0 &  0 \\
      Stardog SHACL~\cite{stardog}          &  0 &  1 &  1 \\
      rudof~\cite{rudof}                  &  1 &  0 &  0 \\
      \bottomrule
    \end{tabular}
    \caption{Validators}
    \label{tab:shacl_tool_usage}
  \end{subtable}
  \caption{Tools for SHACL Extraction and Validation}
\end{table}

\begin{figure}
  \centering
  \includegraphics[width=0.65\linewidth]{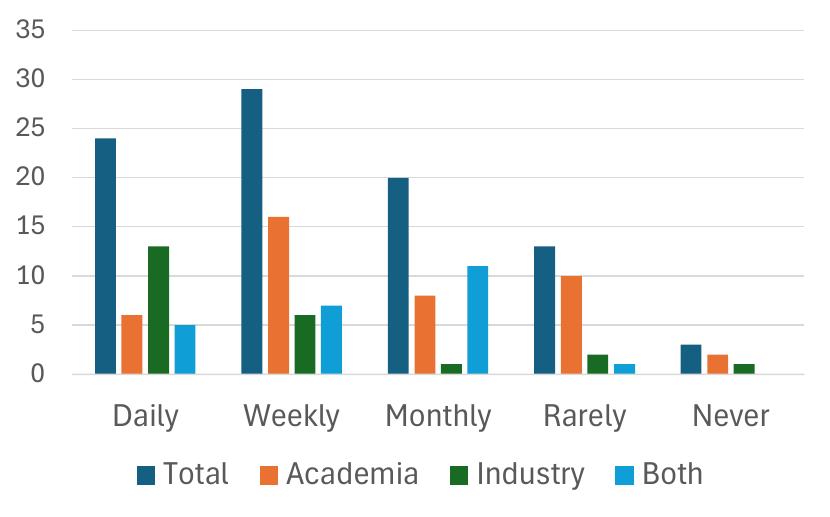}
  \caption{Validation Usage Frequency }
  \label{fig:usage_frequency_background}
\end{figure}

This diverse tooling supports a spectrum of validation
intensities. Notably, validation is \emph{more frequently integrated
  into operational workflows in industry}, with daily or weekly use
being common, compared to more varied or sporadic application in
academic settings (Figure~\ref{fig:usage_frequency_background}). Such
regular usage in industry naturally demands more from both the
languages and the tools. This is also reflected in \emph{how deeply
  language features are utilized}: high-frequency SHACL users, often
in the industry context, are significantly more likely to employ
advanced\footnote{\url{https://www.w3.org/TR/shacl-af/}} capabilities
like SHACL Rules and Functions
(Table~\ref{tab:advanced_shacl_usage}). These advanced features are
not part of the W3C SHACL Recommendation, but are a collection of
useful extensioins of SHACL. This reliance on the advanvced features
indicates that standard features alone can be insufficient in some
data engineering contexts, pushing users towards more expressive
features.

\begin{table}
  \footnotesize
  \begin{subtable}[t]{0.48\textwidth}
    \footnotesize
    \centering
    \begin{tabular}{lrrr}
      \toprule
      Advanced Feature & Acad. & Ind. & Both \\
      \midrule
      Expression constraints & 7 & 4 & 5 \\
      Node expressions       & 9 & 6 & 6 \\
      SHACL functions        & 8 & 10 & 4 \\
      SHACL rules            & 13 & 13 & 10 \\
      \bottomrule
    \end{tabular}
    \caption{Advanced SHACL Feature Usage}
    \label{tab:advanced_shacl_usage}
  \end{subtable}
  \begin{subtable}[t]{0.48\textwidth}
    \footnotesize
    \centering
    \begin{tabular}{lrrr} 
      \toprule
      Graph Size & Acad. & Ind. & Both \\
      \midrule
      1 to 100K triples       & 22 & 7 & 14 \\
      100K to 1M triples      & 8  & 6 & 3  \\
      1M to 100M triples      & 8  & 7 & 4  \\
      More than 100M triples  & 4  & 3 & 3  \\
      \bottomrule
    \end{tabular}
    \caption{Graph Size at Validation Time}
    \label{tab:graph_size_validation}
  \end{subtable}
  \caption{Operational Context}
  \label{tab:operational} 
\end{table}

These practices are applied across varied dataset sizes
(Table~\ref{tab:graph_size_validation}), from manageable academic
datasets to the very large knowledge graphs increasingly found in
industry settings. The interplay of frequent validation cycles, the
need for advanced features, and the challenge posed by large datasets
collectively defines the demanding operational context that current
and future validation technologies should address.

In essence, the way validation is practiced, particularly the higher
frequency, greater use of advanced features, and often larger data
scales in industry, shapes the demands on both languages and
tools. This operational context reveals a \emph{need for robust,
  expressive, and scalable validation solutions}, supported by a
diverse tooling ecosystem that caters to varying levels of complexity
and integration requirements.

\section{Current Challenges and Future Opportunities}
\label{sec:challenges}
Charting the future trajectory for RDF validation requires a clear
understanding of its current standing within the community. Therefore,
we explore identified limitations and desired enhancements to pinpoint
key opportunities for advancement and research.

To get a more complete picture, we start with analizing the perceived
advantages of current validation technologies
(Table~\ref{tab:advantages_limits}). Two primary themes emerge; first,
respondents frequently mentioned the \emph{strong validation
  capabilities} and \emph{flexibility} offered for defining
constraints, indicating general satisfaction with the core expressive
power and design principles of languages like SHACL and ShEx. Second,
the benefit of integration within the wider \emph{RDF ecosystem} and
compatibility with existing tooling was often highlighted,
underscoring the practical value derived from standardization efforts.

\begin{table}
  \label{tab:advantages_issues_combined} 
  \footnotesize
  \begin{subtable}[t]{0.48\textwidth} 
    \centering
\begin{tabular}{@{}p{3.4cm}ccc@{}}
\toprule
\textbf{Perceived Advantage} & \textbf{Ac.} & \textbf{Ind.} & \textbf{Both} \\
\midrule
Compatibility with RDF & 34 & 22 & 20 \\
Strong validation capabilities & 21 & 17 & 20 \\
Flexible constraints & 23 & 13 & 17 \\
Integration with existing tools & 14 & 10 & 12 \\
Ease of use & 16 & 10 & 7 \\
Advanced constraints & 9 & 6 & 13 \\
Documentation quality & 12 & 6 & 7 \\
Community support & 13 & 6 & 5 \\
Comprehensive documentation & 5 & 4 & 2 \\
Performance and scalability & 3 & 3 & 4 \\
\bottomrule
\end{tabular}
  \end{subtable}
  \hfill 
  \begin{subtable}[t]{0.49\textwidth} 
    \centering
\begin{tabular}{@{}p{3.2cm}ccc@{}}
\toprule
\textbf{Issue / Limitation} & \textbf{Ac.} & \textbf{Ind.} & \textbf{Both} \\
\midrule
Documentation and tutorials & 22 & 8 & 11 \\
Community and tool support & 20 & 11 & 8 \\
Performance and scalability & 17 & 9 & 11 \\
Lack of features and functionality & 14 & 11 & 8 \\
Limited constraint types & 11 & 11 & 10 \\
No major challenges & 8 & 5 & 10 \\
Interoperability & 12 & 4 & 0 \\
Usability & 2 & 0 & 3 \\
Lack of implementations & 1 & 2 & 1 \\
Cumbersome syntax & 0 & 1 & 2 \\
\bottomrule
\end{tabular}
  \end{subtable}
  \caption{Perceived Advantages and Reported Issues with Validation Technologies}
  \label{tab:advantages_limits} 
\end{table}

Complementing these advantages, the investigation into desired
features and current limitations reveals key areas needing attention
by the community (Table~\ref{tab:advantages_limits}). Significant
hurdles to wider adoption and more effective use appear to be the
current state of \emph{support, usability, and documentation}. Survey
respondents frequently cited inadequate documentation and tutorials,
alongside gaps in comprehensive tool support, as prominent
challenges. The absence of foundational learning resources, such as an
official W3C SHACL Primer comparable to those for other RDF standards,
was specifically highlighted as a notable shortcoming.

\begin{figure}
    \centering 

    \begin{subfigure}[b]{0.48\linewidth}
        \centering
        \includegraphics[width=\linewidth]{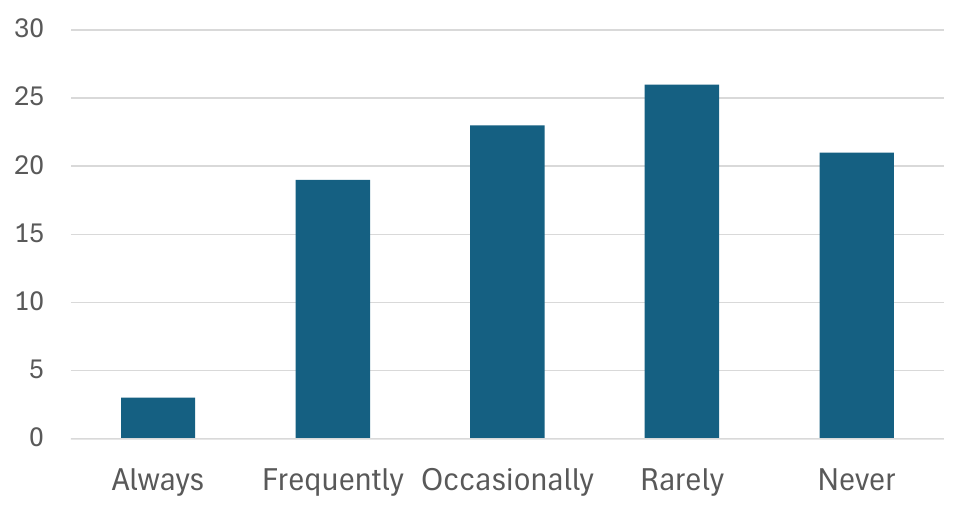}
        \caption{Frequency of Use}
        \label{fig:shaclsparql_freq}
    \end{subfigure}
    \begin{subfigure}[b]{0.48\linewidth}
        \centering
        \includegraphics[width=\linewidth]{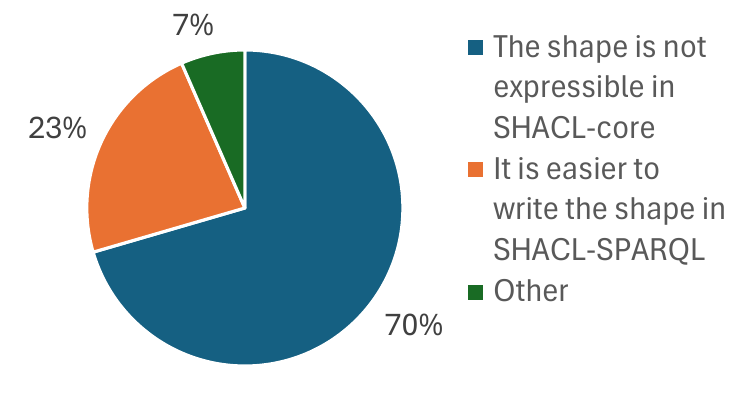}
        \caption{Motivation for Use}
        \label{fig:shaclsparql_why}
    \end{subfigure}

    \caption{SHACL-SPARQL Usage} 
    \label{fig:shaclsparql} 
\end{figure}

\begin{figure}
  \centering 

  \begin{subfigure}[b]{0.46\linewidth}
    \centering
    \includegraphics[width=\linewidth]{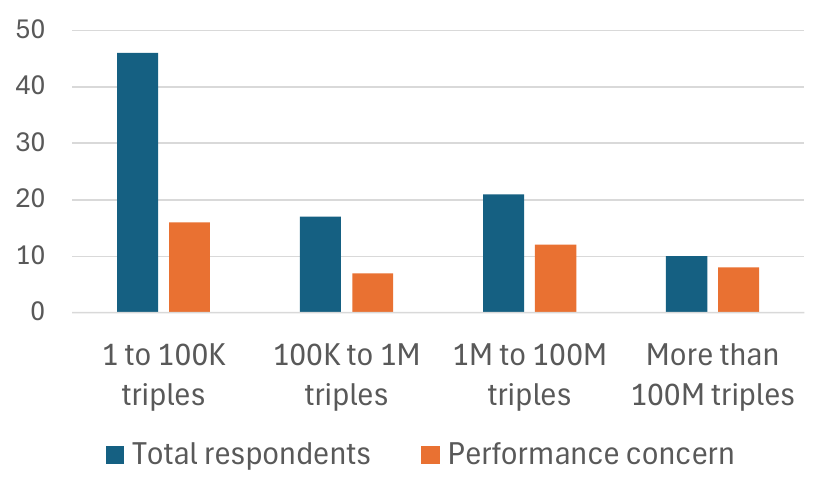}
    \caption{Graph Size vs. Performance Concerns}
    \label{fig:perf_size}
  \end{subfigure}
  \begin{subfigure}[b]{0.5\linewidth}
    \centering
    \includegraphics[width=\linewidth]{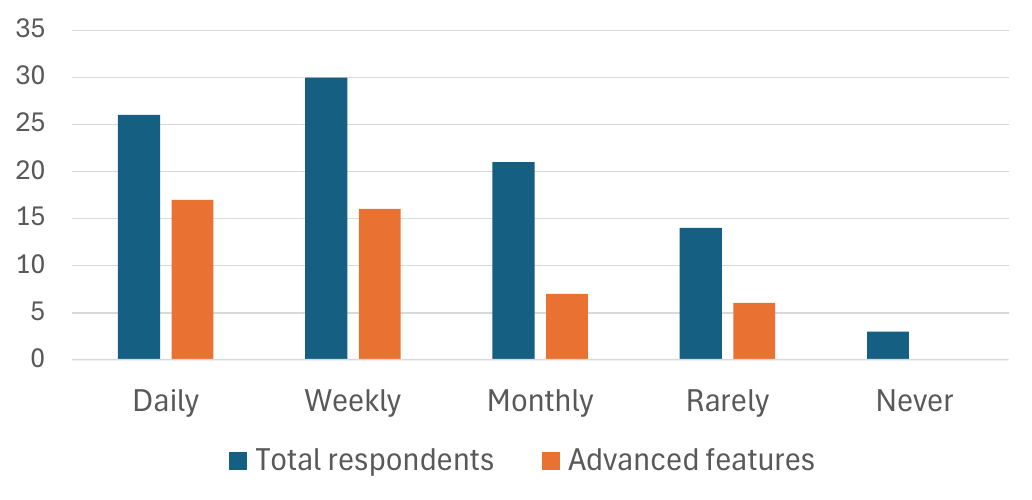}
    \caption{Frequency of Use vs. Advanced Features}
    \label{fig:freq_adv}
  \end{subfigure}

  \caption{Advanced Analysis} 
  \label{fig:advanced_analysis} 
\end{figure}

Furthermore, two critical technical challenges emerged from the
survey: \emph{ensuring adequate performance}, particularly for large
datasets, and addressing perceived \emph{gaps in the expressiveness of
  the core validation languages}. Scalability remains a pressing
concern, with performance bottlenecks frequently reported by users
working with large graphs (Figure~\ref{fig:perf_size}).
While some initial research has moved in the direction of addressing these issues~\cite{shaclsql, travshacl}, the survey indicates that significant challenges remain.
Simultaneously, the survey
indicates that the expressive power of core SHACL is often found
insufficient for complex validation
needs. Figure~\ref{fig:shaclsparql_freq} shows that users often resort
to
SHACL-SPARQL\footnote{\url{https://www.w3.org/TR/shacl/\#sparql-constraints}},
which allows for writing SPARQL-based constraints beyond the standard
components. Moreover, Figure~\ref{fig:shaclsparql_why} shows the
primary reason is the increased expressiveness. Furthermore,
Figure~\ref{fig:freq_adv} shows that frequent users of SHACL, also
tend to more frequently rely on non-standard SHACL Advanced
Features\footnote{\url{https://www.w3.org/TR/shacl-af/}}.

Additionally, specifically desired language extensions, such as robust
support for \emph{validating named graphs}, were also voiced in
free-text answers. Consequently, the call for better tooling,
specifically for SHACL-SPARQL and these Advanced Features, not only
underscores the general tool support gap but also the community's need
for mature implementations of these more complex and expressive
features.

\begin{table}
  \centering
  \begin{subtable}{1.0\textwidth}
      \small    
    \centering
    \begin{tabular}{lr}
      \toprule
      \textbf{Statement} & \textbf{Count} \\
      \midrule
      5 - It is too detailed          & 1              \\
      4 - It is sufficient but too verbose & 7              \\
      3 - It is adequate              & 40             \\
      2 - It needs to be more informative            & 25             \\
      1 - It requires significant enhancement & 11             \\
      \bottomrule
    \end{tabular}
    \caption{Validation Report Usefulness}
    \label{tab:q19_validation_report_sub}
  \end{subtable}
  
  \begin{subtable}{1.0\textwidth}
    \small
    \centering
    \begin{tabular}{lrrr}
    \toprule
    \textbf{Desired Feature} & \textbf{Acad.} & \textbf{Ind.} & \textbf{Both} \\
    \midrule
    Support for RDF 1.2        & 13 & 18 & 12 \\
    Support for recursive shapes & 15 & 12 & 9  \\
    Support for Property Graphs  & 12 & 5  & 5  \\
    Other                        & 4  & 4  & 4  \\
    None                         & 4  & 1  & 2  \\
    \bottomrule
  \end{tabular}
  \caption{Desired Future Language Features}
  \label{tab:desired_features_sub}
\end{subtable}
\caption{User Feedback on Validation Reports and Desired Language Features}
  \label{tab:combined_feedback} 
\end{table}

Table~\ref{tab:desired_features_sub} shows specific areas for future
development. A clear call from the community is for these languages to
evolve in alignment with broader RDF advancements and to address
current functional gaps. This is evidenced by strong interest in
\emph{support for RDF 1.2/RDF-Star}, ensuring validation tools can
handle the latest RDF advancements. Another prominent request is the
high demand for \emph{explicit support for recursive shape constraints
  in SHACL}, where we highlight the strong reaction from respondents
with a background in industry. Given that recursion is a capability in
ShEx but remains undefined in the SHACL specification, this signals a
need for standardized, comprehensive feature sets, a topic noted for
consideration by the W3C Data Shapes working
group~\cite{datashapes_wg}.

Beyond language features, \emph{improving the usability of validation
  reports} also emerged as an
opportunity. Table~\ref{tab:q19_validation_report_sub} indicates that
while SHACL validation reports are generally satisfactory, there is
considerable scope for enhancing their usability, potentially through
more informative error details or better integration with debugging
workflows.

\begin{figure}
    \centering
    \includegraphics[width=0.8\linewidth]{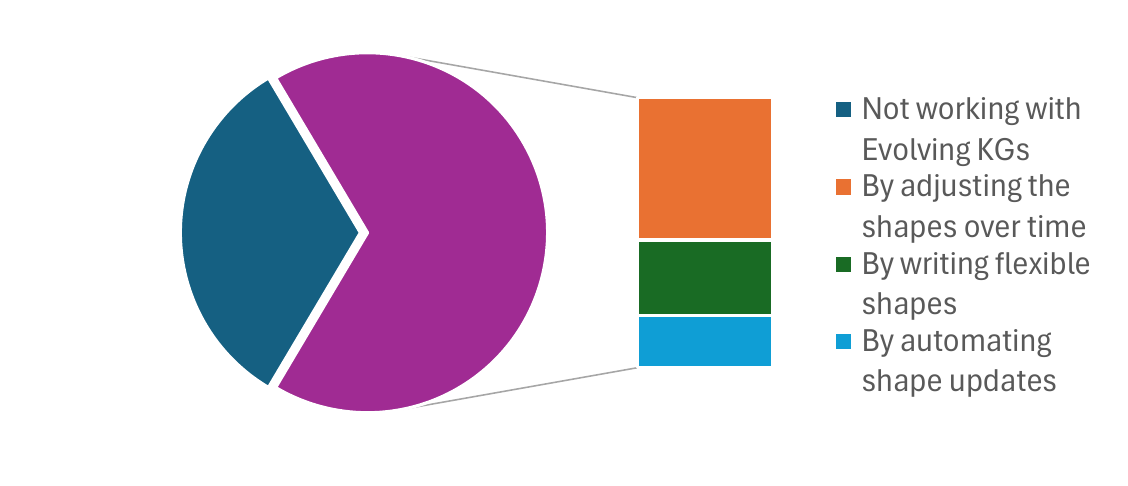}
    \caption{Methods for Validating Evolving KGs}
    \label{fig:evolving}
\end{figure}

Finally, the survey touched upon the increasingly relevant challenge
of managing evolving knowledge graphs~\cite{tgdk_evolving}. A
significant majority (67\%) of respondents reported working with
evolving graph data in their validation workflows
(Figure~\ref{fig:evolving}). Many indicated that changes to the graph
often necessitate manual adjustments to their shape constraints,
highlighting a need for more robust methods or tooling to manage the
co-evolution of data and constraints in dynamic environments. This
presents a key area for future research and innovation.

In summary, while RDF validation technologies like SHACL and ShEx are
valued for their core capabilities and ecosystem integration, our
survey indicates paths for enhancement crucial for both practitioners
and researchers. Firstly, \emph{improving the support infrastructure
  and usability} remains vital for effective adoption. This includes
better documentation and tutorials (e.g., official primers), more
mature tooling, and enhanced validation report utility to aid
debugging and understanding of existing features.

\section{Interactive Data Exploration}
\label{sec:sql}

In our commitment to open science and fostering community engagement,
this section details the methodology behind our data analysis and
introduces the survey dataset as an interactive resource. All figures,
tables, and quantitative analyses presented in this paper are the
direct result of SQL queries executed against a structured and
publicly available version of our survey data.

\subsection{From Raw Responses to Queryable Data}

The journey from the initial survey responses, collected via Google
Forms, to a queryable dataset involved a systematic preparation
process. Firstly, predefined answer options for multiple-choice
questions were assigned consistent codes (e.g., the tool ``Apache Jena
SHACL'' as an answer to Q11 was coded as ``apache''). Secondly,
free-text answers were manually reviewed. Where responses clearly
corresponded to existing coded options, they were mapped
accordingly. For frequently occurring and distinct free-text answers
relevant to our analysis, new codes were introduced. This curated and
coded dataset was then loaded into a
DuckDB\footnote{\url{https://duckdb.org/}} database. The complete
dataset, including the raw CSV files, and the DuckDB database file are
publicly
available\footnote{\url{https://github.com/dmki-tuwien/rdf-validation-community-survey/blob/main/README.md}}.

\subsection{Navigating the Survey Data: A Glimpse into the Schema}

The database is structured to facilitate our analysis through
querying. The primary tables are:
\begin{description}
\item[\texttt{answers}] The central table where each row corresponds
  to a unique respondent. Columns represent individual questions
  (e.g., \texttt{q01} for professional background, \texttt{q11} for
  SHACL tools used). For questions allowing multiple selections,
  answers are stored as arrays of codes.
\item[\texttt{answers\_historical}] This table contains responses from
  the 2022 survey~\cite{rabbani2022shacl}, which had a smaller
  scope. This allows for longitudinal comparisons for questions that
  were part of both survey editions.
\item[\texttt{question\_lt}] A lookup table that maps the question
  IDs (used as column names in \texttt{answers}) to their full,
  textual descriptions (e.g., ``What is your professional
  background?'').
\item[\texttt{qXX\_lt}] A series of lookup tables, one for each
  question \texttt{qXX} (e.g., \texttt{q01\_lt}). These tables
  translate the internal codes used for answers (e.g., ``apache'' for
  Q11) back to their descriptive labels (e.g., ``Apache Jena SHACL'').
\end{description}
This relational structure allows for flexible data aggregation,
filtering, and joining to extract insights.

\subsection{Illustrative Queries}
The quantitative results presented throughout this paper were derived
using SQL queries. For instance, to generate the data for
Figure~\ref{fig:professional_background} (Professional Background of
respondents), the following query was used:
{\footnotesize
\begin{verbatim}
WITH current_answer_counts AS (
    SELECT q01, COUNT(respondent_id) AS current_respondent_count
    FROM answers
    GROUP BY q01 ),
historical_answer_counts AS (
    SELECT q01, COUNT(*) AS historical_respondent_count 
    FROM answers_historical
    GROUP BY q01 )
SELECT qlt.label AS "Background",
    COALESCE(cac.current_respondent_count, 0) AS "Count 2025",
    COALESCE(hac.historical_respondent_count, 0) AS "Count 2022"
FROM q01_lt qlt 
LEFT JOIN current_answer_counts cac ON qlt.identifier = cac.q01
LEFT JOIN historical_answer_counts hac ON qlt.identifier = hac.q01
ORDER BY "Count 2025";
\end{verbatim}
}
This query counts respondents based on their professional background
(Q01), joining with the \texttt{q01\_lt} table to display
human-readable labels. The results of this query are then visualized
in Figure~\ref{fig:professional_background}.

The true power of this open dataset lies in its potential for novel
explorations, allowing for deeper dives into the characteristics and
needs of specific user segments. For instance, one might want to
understand the ``expressiveness chasm'' more acutely by focusing on
SHACL power users. We can pose a research question such as: "What are
the common professional backgrounds, primary motivations for using
SHACL-SPARQL, and most desired SHACL language extensions among those
who not only use SHACL-SPARQL `Frequently' or `Always' but also
specifically employ `SHACL Rules'?"

To answer this, a SQL query can be constructed to first filter our
answers table for respondents matching this precise profile (heavy
SHACL-SPARQL users via Q17 and SHACL Rules users via Q20). For this
identified segment, the query then aggregates their responses
regarding professional background (Q01), their stated reasons for
using SHACL-SPARQL (Q18), and the future SHACL language features they
desire (Q21). The provided example query demonstrates exactly this,
aiming to count how many users within this advanced segment share
particular combinations of these characteristics. We can formulate
this as follows:

{\footnotesize
\begin{verbatim}
WITH TargetUserSegment AS (
    -- Identify respondents who are heavy SHACL-SPARQL users (Q17)
    -- AND use SHACL Rules (Q20).
    SELECT a.respondent_id,
        a.q01, -- background
        a.q18, -- SHACL-SPARQL usage motivation
        a.q21  -- desired SHACL extensions
    FROM answers a
    JOIN q17_lt q17lt 
    ON a.q17 = q17lt.identifier -- SHACL-SPARQL usage frequency
    WHERE q17lt.label IN ('Always', 'Frequently') -- heavy usage
        AND EXISTS ( -- Check if they use SHACL rules
            SELECT 1 FROM UNNEST(a.q20) AS u20(code)
            JOIN q20_lt ON q20_lt.identifier = u20.code
            WHERE q20_lt.label = 'SHACL rules' ) ),
SegmentNeedsAndMotivations AS (
-- For the target user segment, unnest their desired extensions (Q21)
-- and join to get labels for their background (Q01) and motivation (Q18).
    SELECT tus.respondent_id,
        q01lt.label AS background_label,
        q18lt.label AS sparql_motivation_label,
        unnested_q21.extension_code
    FROM TargetUserSegment tus
    JOIN q01_lt q01lt ON tus.q01 = q01lt.identifier 
    JOIN q18_lt q18lt ON tus.q18 = q18lt.identifier,
    UNNEST(tus.q21) AS unnested_q21(extension_code) 
)
-- Count unique respondents within the segment for combinations of
-- prof. backgr., SHACL-SPARQL motivation, and desired SHACL extension.
SELECT snm.background_label AS "Prof. Backgr. (Q01)",
    snm.sparql_motivation_label AS "SHACL-SPARQL Motive (Q18)",
    q21lt.label AS "SHACL Extension (Q21)",
    COUNT(DISTINCT snm.respondent_id) AS "Respondent Count"
FROM SegmentNeedsAndMotivations snm
JOIN q21_lt q21lt ON snm.extension_code = q21lt.identifier
GROUP BY "Prof. Backgr. (Q01)",
    "SHACL-SPARQL Motive (Q18)",
    "SHACL Extension (Q21)"
-- Show only combinations with actual respondents
HAVING COUNT(DISTINCT snm.respondent_id) > 0 
-- Show most common combinations first
ORDER BY "Respondent Count" DESC;
\end{verbatim}
}
The results from this query are shown in
Table~\ref{tab:advanced_shacl_user_profile}. It illustrates the
dataset's potential to highlight patterns even within niche
subgroups. In the example, we observe that for users employing both
frequent SHACL-SPARQL and SHACL Rules, the predominant motivation
appears to be the perceived limitations of SHACL-core, and their
desired extensions frequently include support for RDF 1.2, recursive
shapes, and property graphs. This demonstrates how the dataset can be
used to connect different facets of user experience for highly
specific cohorts.

\begin{table}[htbp]
\centering
\resizebox{\textwidth}{!}{%
\begin{tabular}{@{}l l l r@{}}
  \toprule
  \textbf{Prof. Backgr. (Q01)} & \textbf{SHACL-SPARQL Motive (Q18)} & \textbf{SHACL Extension (Q21)} & \textbf{Count} \\
  \midrule
  Academia & The shape is not expressible in SHACL-core & Support for property graphs & 3 \\
  Academia & The shape is not expressible in SHACL-core & Support for RDF 1.2 & 2 \\
  Academia & The shape is not expressible in SHACL-core & Support for recursive shapes & 2 \\
  Both & The shape is not expressible in SHACL-core & Support for RDF 1.2 & 2 \\
  Both & The shape is not expressible in SHACL-core & Support for recursive shapes & 2 \\
  Industry & The shape is not expressible in SHACL-core & Support for RDF 1.2 & 2 \\
  Industry & The shape is not expressible in SHACL-core & Support for recursive shapes & 2 \\
  Academia & The shape is not expressible in SHACL-core & None & 1 \\
  Both & The shape is not expressible in SHACL-core & Other & 1 \\
  Both & It is easier to write & Support for recursive shapes & 1 \\
  Both & Other & Support for RDF 1.2 & 1 \\
  Both & Other & Support for property graphs & 1 \\
  Industry & The shape is not expressible in SHACL-core & Other & 1 \\
  Industry & The shape is not expressible in SHACL-core & Support for property graphs & 1 \\
  \bottomrule
\end{tabular}%
}
\caption{Profile of Advanced SHACL Users}
\label{tab:advanced_shacl_user_profile}
\end{table}

While the example query showcases the dataset's ability to identify
and profile niche user segments, the small respondent counts in such
granular analyses mean these specific findings are more indicative
than statistically definitive. The dataset's true value here is in
demonstrating its potential as a starting point for diverse research
inquiries, which could then be explored further, perhaps with
qualitative methods.

\subsection{Your Turn: Access the Data and Tools}
To immediately explore the data and begin deriving your own insights,
we invite you to use the in-browser queryable instance of our survey
data available on our website. You can write your SQL queries directly
in the browser and see results
interactively\footnote{\url{https://dmki-tuwien.github.io/rdf-validation-community-survey/databaseshell}}. For
those who prefer to work locally or integrate the data into other
analysis environments, the complete dataset, including the DuckDB
file, is also available for download. DuckDB is a lightweight,
file-based SQL database that requires no complex setup, making it easy
to query using various tools.

We encourage you to explore the data, replicate our findings,
challenge our interpretations, and, most importantly, \textbf{derive
  your own insights!}

\section{Conclusions}
\label{sec:conclusion}

In this paper, we presented an analysis of the RDF validation
landscape, drawing upon our recent community survey. Our findings
offer a contemporary snapshot of how SHACL and ShEx are adopted, the
methods used for creating shapes and executing validation, and the
prevailing challenges and needs. The survey highlights the maturation
of RDF validation, particularly the widespread experience with SHACL
across diverse domains and user backgrounds, including a notable
recent growth in academia alongside persistent industry
engagement. While validation is widely perceived to enhance data
quality, and automated shape generation from instance data is gaining
traction, significant challenges persist. Our analysis underscores a
consistent demand for improved usability (especially better
documentation and more intuitive tooling), enhanced performance at
scale, and greater language expressiveness to meet complex real-world
requirements.

To foster further exploration of these nuances and encourage
community-driven insights, we have made our survey dataset publicly
available. This data, which underpins all quantitative results in this
paper, is provided as a queryable DuckDB SQL database. While an RDF
representation of such survey data might be an ideal for direct
integration within the RDF ecosystems (a step that would involve an
additional mapping effort) we chose DuckDB for its immediate
practicality. It offers an efficient, integrated, in-memory
environment for readily using the cleaned data, and notably supports
in-browser querying through WASM, facilitating broad
accessibility. This approach streamlined our analysis, and we hope it
empowers others; indeed, a direct, DuckDB-like interactive querying
experience tailored for RDF datasets could be a valuable avenue for
future tooling development within our field.

Building on the insights from our study, our work underscores several
key directions for the community, encompassing researchers, tool
developers, standards bodies, and practitioners. There is a clear need
to improve performance for large-scale validation and to advance shape
creation tooling, making it more sophisticated and
user-friendly. Extending the expressiveness of SHACL, for instance, by
standardizing support for recursive shapes and comprehensive
validation of named graphs, remains a demand. Furthermore, enhancing
both documentation (perhaps with official primers) and the utility of
validation reporting is vital for wider adoption and effective
use. Supporting validation in evolving knowledge graphs also emerges
as a critical area for future research and development, focusing on
strategies for the co-evolution of data and constraints. Addressing
these areas will be crucial for the continued growth and effectiveness
of RDF validation in ensuring data quality and the trustworthiness of
knowledge graphs

\bibliographystyle{plain}
\bibliography{references}

\end{document}